\documentclass[twocolumn,pra]{revtex4-1}
\usepackage[T1]{fontenc}
\usepackage[utf8]{inputenc}
\usepackage{amsmath,amssymb,bbold}
\usepackage{color}
\usepackage{graphicx}
\usepackage{amssymb}
\begin{document}
\title{Spin Orbit Coupling in Graphene Induced by Heavy Adatoms with Electrons in the Outer-Shell $p$ Orbitals. 
}
\author{ Luis Brey  }
\email{Electronic address: brey@icmm.csic.es}
\affiliation{Departamento de Teor\'{\i}a y Simulaci\'on de Materiales, Instituto de Ciencia de Materiales de Madrid, CSIC, 28049 Cantoblanco, Spain}

\date{\today}

\begin{abstract}
Many of the exotic properties proposed to occur in graphene rely on the possibility of increasing the spin orbit coupling (SOC). By combining analytical and numerical tight binding calculations, in this work we study
the SOC induced by heavy adatoms with active electrons living in $p$ orbitals.  Depending on the position of the adatoms on graphene different kinds of SOC appear.
Adatoms located in hollow position induce spin conserving intrinsic like SOC whereas a random distribution  of adatoms induces a  spin flipping Rashba like SOC.  The induced SOC  is  linearly proportional to the adatoms concentration, indicating the inexistent interference effects between different adatoms.  By computing the Hall conductivity we have proved the stability of  the topological quantum Hall phases
created by the adatoms against inhomogeneous spin orbit coupling .   For the case of Pb adatoms, we find that a concentration of 0.1 adatom per carbon atom generates SOC's of the order of $\sim$40$meV$.

\end{abstract}

\maketitle
\section{Introduction}

The research on graphene, a two-dimensional crystal of carbon atoms, has driven  to  the discover of  a large number of interesting electrical, magnetic, mechanical and optical properties\cite{Guinea_2009,Katsnelson-book}.
The small atomic number of carbon implies that electrical carriers in graphene have a extremely weak spin-orbit (SO)  coupling\cite{Huertas-2006,Min-2006}. This property  in combination with the large graphene electron mobility 
makes graphene a very good candidate for using in spintronics\cite{Dlubak-2012,Chico-2015a}.

On the other hand some proposasl of exotic topological phases in graphene rely on the possibility of increasing  the SOC. Because of  the graphene  lattice symmetry, there are two types of spin orbit coupling in graphene, {\it intrinsic}-like, where the ${\hat z}$-component of the electron spin is a good quantum number and ${\it Rashba}$-like which mixes spins and  appears in absence of mirror symmetry\cite{Kane-2005}. 
In graphene, the intrinsic SOC opens a gap and the system becomes a quantum spin Hall insulator, with gapless edge states able to transport spin and charge\cite{ Hasan-2010,Qi-2011}.  Non-trivial topological phases may also occur  in multilayer graphene \cite{Prada_2011}.  Quantum anomalous Hall effect was predicted to occur in  bilayer and monolayer graphene in presence of  Rashba SOC   and  an exchange field or magnetic impurities\cite{Qiao_2010,Tse_2011}.
The experimental realization of  these topological phases requires a large SOC, and therefore there is a big interest\cite{CastroNeto-2009,Abdelouahed-2010,Chico-2009,Gosalbez-2011, Balakrishnan-2013,Calleja-2015,Avsar-2014} in increasing the SOC and clear the way to the study of exotic topological phases in graphene. 
Experimental reports on enhancement of SOC in graphene by weak hydrogenation\cite{Balakrishnan-2013}, gold hybridization\cite{Marchenko-2012} or proximity with  WS$_2$\cite{Avsar-2014}, indicate that it is possible to increase the SOC in almost three orders of magnitude.  Recently, it has been reported that graphene grown on Cu shows a SO splitting around 20 meV\cite{Balakrishnan-2014}. Intercalation of Au atoms in graphene grown on Ni  produces a SO splitting of near 100meV\cite{Marchenko-2012}.  Similarly, the intercalation of Pb atoms in graphene grown on a iridium  substrate seems to produce a giant SOC\cite{Calleja-2015}. Theoretically, it has been proposed that heavy adatoms with partially filled $p$-shells, deposited on symmetric positions of the graphene lattice,  could induce large intrinsic SOC\cite{weeks_2011}.

\begin{figure}[htbp]
\includegraphics[width=\columnwidth,clip]{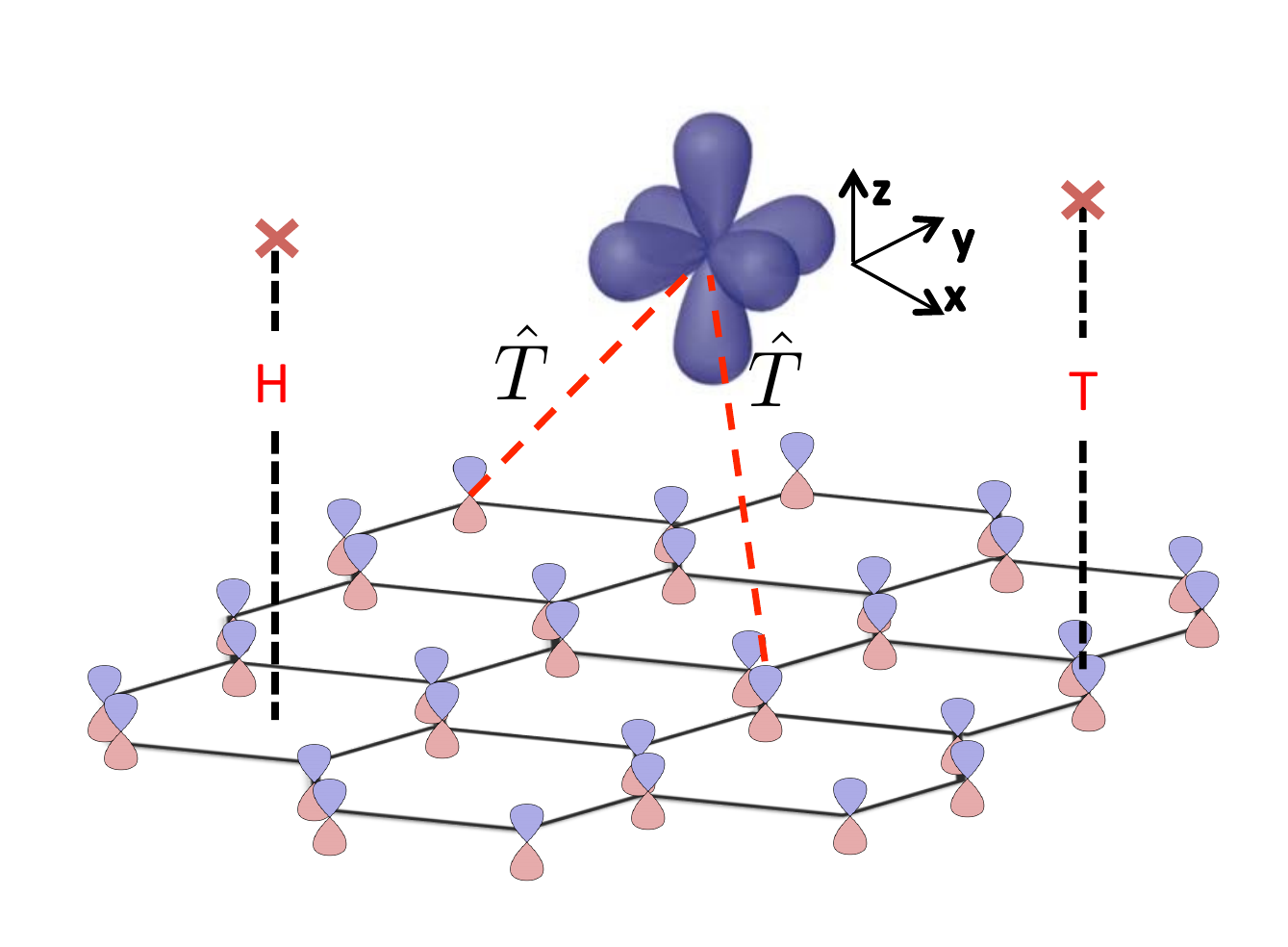}
\caption{(Color online)Schematic representation of the effective hopping between  carbon atoms induced by an adatom with
active electrons in outer $p$-shells. Vertical, H and T, lines indicate the hollow and top positions respectively}                                     
\label{Figure1}
\end{figure}

In this work we study the SOC induced by heavy adatoms with active electrons living in p-orbitals, in particular we consider Pb atoms, that have been proved  to induce large SO effects in 
graphene\cite{Calleja-2015}.

The physical picture is the following, the tunneling of an electron between two carbon atoms through the  adatoms $p$-orbitals opens new channels for hopping in graphene.
The SOC  between the adatom $p$ orbitals makes that  
the new  tunneling channels can conserve the spin inducing a intrinsic SOC  or can flip the spin inducing a Rashba like SOC.

By combining analytical calculations, perturbation theory and tight-binding based numerical simulations, we study the type of SO coupling induced by adatoms residing in different positions of the graphene unit cell. In addition, we study how a finite density of adatoms, 
in different distributions, affects the induced SO couplings.
The main conclusions of our work are the following, 
\par\noindent i) adatoms located in hollow positions, see Fig.\ref{Figure1}, induce intrinsic SOC. 
A finite density of adatoms in hollow positions opens an energy gap at the 
Dirac points, that increases linearly with the adatom concentration. This gapped phase is a quantum spin Hall state. 
The simulations indicate that, even for high adatom coverage,  there are not interference effects between the adatoms
and the gap only depends on adatom density. In the case of Pb atoms we find gaps of the order of 50meV
 for a concentration of 0.1 adatom per carbon.

\par\noindent ii) for adatoms placed in top positions, see Fig.\ref{Figure1}, the tunneling from graphene 
to the adatom and back, induces a Rashba like spin flip  hopping between the underneath C atom and its first neighbors
and an intrinsic like spin conserving second neighbors tunneling between the carbons surrounding the underneath carbon atom. 
The intrinsic like SOC induced by adatoms in  top positions has opposite sign than the induced by  adatoms in 
hollow geometry.
The Rashba SOC has the same sign independently of the sublattice of the underneath carbon.

\par\noindent iii) A finite density of adatoms randomly distributed on graphene, induces a finite Rashba SOC linearly 
proportional to the density  of adatoms. 
For a random distribution of adatoms, the resulting intrinsic like SOC vanishes, because contributions from different
locations of the adatoms have opposite signs.
Similar results are
obtained when the adatoms form an array  commensurate with a large graphene supercell.
By computing the Hall conductivity, we have obtained that  a random distribution 
of adatoms on graphene in presence of  an exchange field, is an anomalous quantum Hall system.
When the adatom is Pb, we obtain that the Rashba SO coupling  can be as larger as 35meV for a   concentration of 0.1 Pb per carbon atom.

The rest of the paper is organized in the following way, in Section II we introduce the graphene and adatom Hamiltonians,
and in Section III we describe the hopping between graphene  carbon atoms and the adatom p-orbitals. 
In Section IV we present the perturbation theory for describing the adatom mediated effective hopping between carbon atoms.  The knowledge of the
effective hopping between carbon atoms  allow us  to obtain, in Section V, analytical expressions for SOC  induced by  adatoms located in top and hollow positions.  Section VI turns to present  tight-binding based numerical simulations for 
studying the effect that a  random or  commensurate distribution of adatoms have on the induced SOC. In Section VII we calculate the topological properties of graphene doped with adatoms. We close the paper with a summary of the results.

\section{Preliminaries.}
\subsection{Graphene Hamiltonian}
In graphene, carbon atoms crystallize in a triangular lattice of primitive translation vectors ${\bf a}$=$(0,a)$ and ${\bf b}$=$(\sqrt{3}/2,1/2)a$, where  $a$=2.46{\AA} is the lattice constant. The positions of the triangular lattices are ${\bf R}_i$. There are two  atoms per unit cell located at positions ${\bf d} _A$=$(0,0)$ and ${\bf d }_B$=$(a/\sqrt{3},0)$, that define sublattices $A$ and $B$ in graphene.
Covalent $sp_2$ bonds between carbon atoms stabilize this honeycomb lattice whereas the tunneling between $p_z$ orbitals is the origin of  the low energy active conduction and valence $\pi$-bands. 
The band structure is rather well described by a tight-binding model with hopping $t \sim 2.7$eV between first neighbors carbon  $p_z$ orbitals,
\begin{equation}
H_0=-t \sum _{<i_A,i_B>,\sigma} \left (|Z,i_A ,\sigma ><Z,i _B,\sigma|
+ H.c. \right)
\label{TB-Graphene}
\end{equation} Here the sum runs over first neighbors pairs and $|Z,i _{\alpha},\sigma> $ represents  the  wavefunction of an electron at position ${\bf R}_i$+$ {\bf d}  _{ \alpha}$ occupying a carbon  $p_z$ orbital with $z$-component of the spin $\sigma$. 
The energy of the  $p_z$ carbon orbital  is chosen as the zero of energies. 

In graphene the intrinsic SOC has a chiral structure of the form, 
\begin{equation}
H_{SO} ^I \! = \!\lambda ^{SO} \!\!\!\! \!\! \! \! \sum _{<i_{\alpha},j _{\alpha}>,\sigma} \! \! \! \!  \left ( i  \sigma (\hat {\bf u}_{i_{\alpha}} \! \times \! \hat {\bf u}_{j_{\alpha}}) _ {z}  |Z,i_{\alpha}  ,\sigma \! >< \! Z,j _{\alpha},\sigma| \!+\!  H.c.\right )
\label{InSOtb}
\end{equation}
where the sum runs over second nearest neighbors carbon atoms and $\hat {\bf u} _{i _\alpha}$ is a unit vector parallel to $\bf R _i$+${\bf d}  _{\alpha}$. Note that the intrinsic SOC conserves spin 
and it is not associated with broken mirror symmetry.
On the contrary, Rashba SOC appears because of broken mirror symmetry, in particular due to the substrate, and 
induces a coupling between first neighbors  with opposite 
spin of the form, 
\begin{equation}
H_{SO} ^R \! =  \! i \lambda ^{R} \!\!\!\! \!\! \! \! \! \!  \sum _{<i_{A},j_{B}>,\sigma ,\sigma '} \! \! \! \! \! \! \left (  ( \boldsymbol \sigma \! 
  \times \! \hat {\bf u}_{i _{A}j_{B}} )_z |Z,i _A,\sigma \!>< \!Z,j_B ,\sigma'| + H.c.\right )
\label{RSOtb}
\end{equation}
here $\hat {\bf u}_  {i_{\alpha}  j _{\beta}} $ is a unit vector parallel to ${\bf R} _j +{\bf d}  _{\beta}-{\bf R} _i -{\bf d} _{\alpha}$ and
 $\boldsymbol  \sigma $ the electron spin  Pauli matrices. 

In absence of SO couplings the conduction and valence bands touch at two inequivalent points of the Brillouin zone
${\bf K}$=$(0,\frac{4 \pi} {3a})$ and ${\bf K}'$=$(0,-\frac{4 \pi} {3a})$ which are the celebrated  Dirac points. Near these points the low energy physics is  described by the Dirac equation
\begin{equation}
H_0=\hbar v_F ( k_x \sigma _0 \otimes  \tau _x + s k_y \sigma _0  \otimes \tau _ y) \, 
\label{Dirac}
\end{equation}
here  the moment $\bf {k}$ is measured with respect the Dirac points, $s$=1 and -1 stands for  ${\bf K}$ and $\bf K '$ respectively, and   $\boldsymbol  \tau $ are the Pauli matrices acting on the spinor defined by the amplitude of the wave function on sublattices $A$ and $B$. In the previous equation $\sigma _0$ represents the unity matrix in the spin sector. The Fermi velocity is related with the hopping trough the relation $\hbar v_F$=$ \frac {\sqrt{3}}2 t a$. In this continuum approximation the SO terms get the form,
\begin{eqnarray}
H_{SO} ^I & = &3 \sqrt {3} \lambda ^{SO} s \,  \sigma _z  \otimes \tau _z \\
\label{SOcontIn}
H_{SO} ^R & =&  \frac 3 2 \lambda  ^R (s \, \sigma _x \otimes \tau _y- \sigma _y \otimes \tau _x) \, .
\label{SOcontRashba}
\end{eqnarray}

\subsection{Adatom Hamiltonian.}
We consider heavy atoms with the active electrons living in  $p$ orbitals. The Hamiltonian describing the electrons in the adatom contains a spin-orbit coupling part and a crystal field $H_{CF}$ term. In the  basis$\{ |p_x \uparrow>,|p_y \uparrow>,|p_z \uparrow>,|p_x \downarrow>,|p_y\downarrow>,|p_z \downarrow>\}$  the Hamiltonian reads, 
\begin{eqnarray}
& &H_{p} =  \Delta _{SO} {\bf L}\cdot {\boldsymbol \sigma } + H_{CF} =  
\nonumber \\ & &
\frac {\Delta _{SO}} 2 \! \left   (  \! \begin{array}{cccccc}
0 &- i &0&0&0& 1\\
i &0&0&0&0&-i\\
0&0&0&- 1&i &0 \\
0&0&- 1&0&i&0\\
0&0&-i &-i &0&0\\
 1&i&0&0&0&0 
\end{array} \! 
 \right ) \!  \! + \! \!  
\left ( \! \begin{array}{cccccc}
\epsilon _x &0 &0&0&0& 0\\
0 &\epsilon _y&0&0&0&0\\
0&0&\epsilon _z&0&0 &0 \\
0&0&0&\epsilon _x&0&0\\
0&0&0 &0 &\epsilon_y&0\\
0&0&0&0&0&\epsilon _z 
\end{array} \! 
\right ) \, \, .
\label{Hp}
\end{eqnarray}
Here $\Delta _{SO}$ is the  spin-orbit coupling parameter, ${\bf L}$  and ${\boldsymbol \sigma}$ are the usual angular momentum and spin operators. Non spherical effects occurring in the geometry produce a crystal field that
splits the energies of the $p$
orbitals. For adatoms deposited on  planar graphene, we expect that $\epsilon_x$=$\epsilon _y \ne \epsilon _z$.

\begin{figure}[htbp]
\includegraphics[width=\columnwidth,clip]{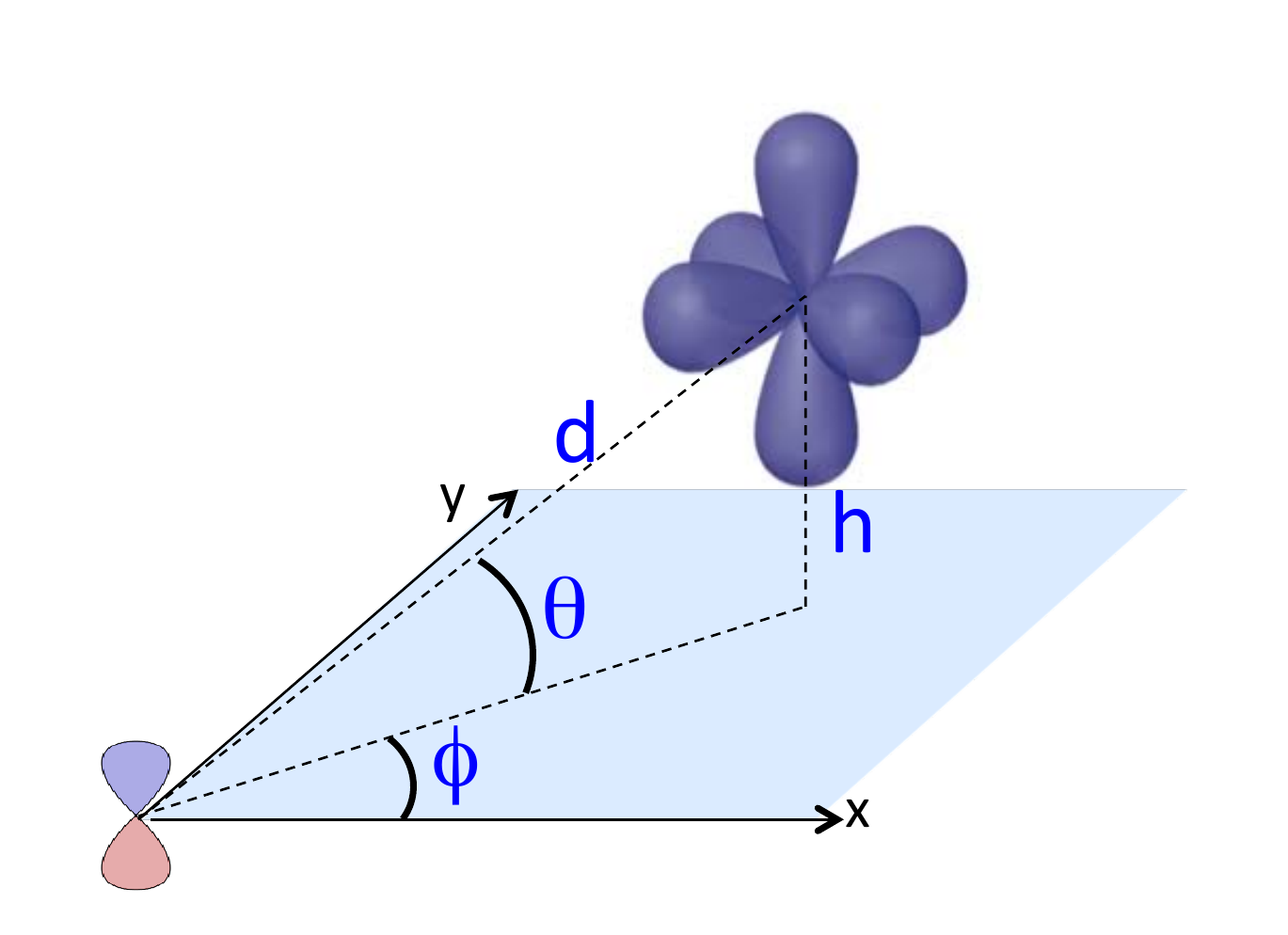}
\caption{(Color online)Geometry of and adatom located at a distance $h$ of the graphene layer. The angles $\theta$ and $\phi$  and the distance $d$  define the spherical coordinates of the adatom with respect to a carbon atom.}                                     
\label{Figure2}
\end{figure}

\section{Tunneling between an  adatom and a carbon $Z$ orbital.}
We consider an adatom placed at position 
${\bf r}$=$(x,y,h)$, where $h$ is the vertical distance between graphene and the adatoms, Fig.\ref{Figure2}. The tunneling amplitudes between an   carbon  orbital located at position
${\bf R}_i$=$(X_i,Y_i,0)$,  and the adatom $p_x$, $p_y$ and $p_z$  orbitals are,
\begin{eqnarray} 
<Z,i,\sigma|\hat T|p_x,\sigma>&=& \frac 1 2 \cos \phi \sin {2 \theta} (V_{pp\sigma} (d)\! -\!V_{pp\pi } (d) ) \nonumber \\ 
<Z,i,\sigma|\hat T|p_y,\sigma>&=& \frac 1 2 \sin \phi \sin {2 \theta} (V_{pp\sigma}(d)\!-\!V_{pp\pi }(d)) \nonumber \\ 
<Z,i,\sigma|\hat T|p_z,\sigma>&=& \cos ^2  \theta  V _{pp\pi}(d)+\sin ^2 \theta V_{pp\sigma}(d)
\label{hoppings}
\end{eqnarray}
where $\theta$=$\tan ^{-1} \frac h {\sqrt{(x-X_i)^2+(y-Y_i)^2}}$, 
$\phi$=$\tan ^{-1} \frac {y-Y_i} {x-X_i}$, and $V_{pp\sigma}$ and $V_{pp\pi}$ are the Slater-Koster hopping parameters between the $p_z$ graphene orbital and the heavy adatom $p$-orbitals.
The hopping parameters depend on the distance $d$=$\sqrt{(x-X_i)^2+(y-Y_i)^2+ h^2}$ between the atoms,  that we parametrize in the form, 
$V_{pp\sigma (\pi)}(d)$=$V_{pp\sigma(\pi)}(d=h)e ^{-\beta (d-h)}$ with $\beta$=3\cite{Morell_2011b} and  $V_{pp\sigma (\pi)}(h)$ obtained from density functional calculations\cite{Calleja-2015}.
Note that  in the tunneling process the carrier spin is conserved.
Because of the symmetry of the $p$-orbitals, the hopping amplitude between a carbon $p_z$-orbital in graphene and the $p_x$ and $p_y$ adatom orbitals, have opposite sign depending weather the adatom is deposited on top or bottom of the graphene sheet. On the contrary, the hopping between $z$-orbitals is independent of the position of the adatom with respect the 
graphene layer.

\section{Tunneling between carbon atoms mediated by an adatom.}
The spin-orbit coupling in the  adatom located at  ${\bf r}$=$(x,y,h)$, allows an extra path for tunneling between two carbon atoms  located at ${\bf R}_i$ and ${\bf R} _j$  with spin $\sigma$ and $\sigma '$ respectively. In second order perturbation theory a single  adatom produces a coupling between the carbon atoms of the form,
\begin{equation}
\gamma _{i\sigma,j\sigma '} = \sum _l \frac {<Z,i,\sigma |\hat T | \, \tilde {p} _l  > < \tilde {p} _l \,  | \hat T  | Z,j,\sigma '>} {\tilde {\epsilon} _l} 
\end{equation}
here $|\tilde p _l>$ and $\tilde {\epsilon} _l$ are the eigenfunctions and eigenvalues of Hamiltonian Eq.\ref{Hp}. 
Because of the form of spin-orbit  coupling in the adatom  outer $p$ shell, the induced hoppings satisfy  the following relations for $i \ne j$
\begin{eqnarray}
&\gamma& _{i\sigma,j-\sigma} =  -\gamma ^*  _{i-\sigma,j\sigma} \nonumber \\
&\gamma& _{i\sigma,j\sigma} =  \gamma ^*  _{i-\sigma,j-\sigma}\nonumber \\
&\gamma &_{i\sigma,j\sigma'} =  \gamma ^*  _{j\sigma ',i\sigma} 
\label{sime}
\end{eqnarray}
and  for $i=j$
\begin{eqnarray}
&\gamma& _{i\sigma,i\sigma} =  \gamma   _{i-\sigma,i-\sigma}\\
&\gamma& _{i\sigma,i-\sigma} = 0
\end{eqnarray}

The adatom gives rise to  two kind of SO assisted tunneling,
\par \noindent
i) spin conserved tunneling  events of the form
\begin{equation}
|Z ,i,\sigma\! > \stackrel {\hat T }\longrightarrow |p_{x(y)} \sigma \! > \stackrel {H_p}\longrightarrow |p _{y(x)} \sigma \! > \stackrel {\hat T} \longrightarrow |Z ,j,\sigma \! >
\label{intrin}
\end{equation}
that are  pure imaginary and change  sign when reversing  spin. Following the standard notation we call it {\it intrinsic} spin-orbit coupling. 
This tunneling amplitude behaves as $\sin ^2 (2 \theta)$, it is zero when the adatom is in the graphene sheet, being independent  on the top or bottom position of the adatom with respect the graphene layer. 
\par \noindent
ii) non-conserving spin processes  of the form 
\begin{eqnarray}  
|&Z& ,i,\sigma \! \! > \stackrel {\hat T }  \longrightarrow |p_z \sigma \!  \! > \stackrel {H_p}\longrightarrow |p _{x(y)} -\sigma\! \! > \stackrel {\hat T} \longrightarrow |Z ,j,-\sigma \!  \! > \nonumber  \\
|&Z &,i,\sigma \! \! > \stackrel {\hat T }  \longrightarrow |p_{x(y)} \sigma \! \! > \stackrel {H_p}\longrightarrow |p _z -\sigma \! \! > \stackrel {\hat T} \longrightarrow |Z ,j,-\sigma\! \! >
\label{Rashba}
\end{eqnarray}
These terms get     origin on the
lack of mirror symmetry in  the hopping between $z$-orbitals and we refer to this  tunneling contribution as 
Rashba SOC. 
This tunneling amplitude behaves as $\sin ( 2\theta)$ and  changes  sign when the adatom is located  on top or on bottom of the graphene layer. 

\section{Effective Hamiltonians for Hollow and Top Positions}
When the adatoms are located  at  high symmetry points of the graphene lattice,  it is possible to write down analytic expressions for the
effect  that the  SO induces  on the  graphene low energy band structure. The procedure consist in 
projecting  the perturbation created by the adatom on the atomic Bloch states at the ${\bf K}$ and ${\bf K '}$ Dirac points,
\begin{eqnarray}
\Psi _{A,s,\sigma } & =& \frac 1 {\sqrt{N}} \sum _i e ^{i s {\bf K}{\bf R}_i} |Z,i_A,\sigma >\nonumber \\
\Psi _{B,s,\sigma} & =& \frac 1 {\sqrt{N}} \sum _i e ^{ i s {\bf K}{\bf R}_i}  |Z,i_B,\sigma >  \, ,
\end{eqnarray}
here $N$ is the number of of unit cells in the crystal.  These Bloch states are the eigenstates of the   Dirac Hamiltonian, Eq.\ref{Dirac}  for ${\bf k}$=0.
We assume that adatoms do not induce coupling between states coming from different Dirac cones. We have checked numerically  this assumption,
provided the adatoms  do not form a periodic array with a reciprocal lattice vector equal to  ${\bf K - \bf K'}$. 

\subsection{Adatom in Hollow Position}
\label{subsec:hollow}
In the hollow geometry the adatom is located on top of the center of an hexagon of the graphene lattice at a height $h$, see Fig.\ref{Figure1}. 
We consider  SO induced tunneling up to third  neighbors,  tunneling between more distant atoms can be neglected because  of the exponential decreasement 
of the tunneling amplitude with the distance. 
Coupling between Bloch wavefunctions of  different sublattices involves first  and third neighbors hopping and gets the form
\begin{equation}
< \Psi _{A,s,\sigma} |V_H| \Psi _{ |B,s,\sigma '} > = \frac 1 N  \sum  _{i_A, j_B}  \gamma _{i _{A} \sigma,j _{B}  \sigma '} e ^{i s {\bf K}({\bf R}_j -{\bf R}_i)}
\label{ABhollow}
\end{equation}
here  $i_A$ ($j_B$)  runs over the vertices, of sublattice $A$ ($B$),  of the hexagon surrounding the adatom.
$V_H$ represents the perturbation created by the adatom in hollow position.

The coupling between Bloch states of the same sublattice involves second neighbors tunneling and gets the form,
\begin{equation}
< \Psi _{A,s,\sigma} |V_H| \Psi _{ |A,s,\sigma '} > = \frac 1 N  \sum  _{i_A \ne j_A}  \gamma _{i _{A} \sigma,j _{A}  \sigma '} e ^{i s {\bf K}({\bf R}_j -{\bf R}_i)}.
\label{AAhollow}
\end{equation}
Similar expression applies for  $< \Psi _{B,s,\sigma} |V_H| \Psi _{ |B,s,\sigma '} > $. The adatom also induces diagonal selfenergies that 
for the adatom in the hollow position are equal for both Dirac points, spin orientation and graphene sublattices. 

In the hollow geometry and for $\sigma'$=$-\sigma $, it is possible to sum the six terms, Eq.\ref{Rashba},  that contribute to  spin flip  effective tunneling, and we 
get
\begin{equation} 
\gamma _{i \sigma,j -\sigma} =- \sigma t_R \left ( e ^{ -i \sigma  \phi _j} - e ^{-i \sigma \phi _i} \right )
\end{equation} 
being $t_R$ a constant that depends on the carbon to adatom tunneling parameters, Eq.\ref{hoppings}.  Using this expression and  relations Eq.\ref{sime}  we obtain that in the hollow position an  adatom with outer shell $p$ orbitals does not induce non-conserving spin tunneling and therefore does not induce Rashba like SOC in graphene.  

For spin conserving SO induced tunneling, the sum of the two processes described  in Eq.\ref{intrin}, gives a hopping,
\begin{equation} 
\gamma _{i \sigma,j \sigma} =i  \sigma t_{so} \sin \left (   \phi _i - \phi _j \right )
\end{equation} 
where $t_{so}$ is a constant  that depends on the distance between the adatom and the graphene sheet.
When introducing this hopping  and applying the symmetries Eq.\ref{sime}, we obtain that the coupling between Bloch states of different sublattices  and same spin cancels identically.  On the contrary the  spin conserving coupling between same sublattice  Bloch functions  gets a finite value that changes sign
when changing spin, sublattice or Dirac cone,
\begin{equation}
< \Psi _{\tau,s,\sigma} |V_H| \Psi _{ \tau,s,\sigma } >= \frac 1 N 3 \sqrt{3}  t_{so} \,  \sigma \, s \, \tau \, .
\label{Intri_H}
\end{equation}
This term has the same form than the Hamiltonian Eq.\ref{SOcontIn} and  we conclude, in agreement with reference \cite{weeks_2011}, that  an heavy   adatom,  with electrical active  $p$-orbital, in a hollow position on top of graphene induces an intrinsic like SOC.

\subsection{Adatom in Top  Position}
In this geometry the adatom is located vertically on top of a carbon atom at a height $h$.
This configuration privileges the sublattice $A$ of the underneath carbon atom. In the top arrangement the carbon  $p_z$  orbital is  orthogonal to the
$p_x$ and $p_y$ orbitals of the adatom  located on top of it. Therefore the processes contributing to first neighbors spin conserving tunneling are
zero by symmetry, Eq.\ref{intrin}.
However, an adatom on top of a carbon of a given sublattice, induces spin conserving tunneling
between carbon atoms of the opposite sublattice,
\begin{equation}
< \! \Psi _{B(A),s,\sigma} |V_{T_{A(B)}}| \Psi _{ B(A),s,\sigma } \!  >= \tau_{A(B)} \frac {3 \sqrt{3}} N   t_{so} \,  \sigma \, s   , 
\label{Intri_T}
\end{equation}
here $V_{T_{A(B)}}$ represents the perturbation created by the adatom on top of atoms belonging to sublattice $A(B)$. 
Therefore adatoms in top positions induce intrinsic-like SO coupling, although it is important to note that
the sign of this conserving tunneling is opposite to the induced by an adatom in hollow position, Eq.\ref{Intri_H}

Because of the symmetry of the orbitals, only one of the mechanisms 
described in Eq.\ref{Rashba} contributes to spin flip tunneling between first neighbors,
\begin{equation} 
\gamma _{i \sigma,j -\sigma} =- \sigma t_R e ^{ -i \sigma  \phi _j} \, ,
\label{Top-in}
\end{equation}  
where $t_R$ is a constant that depends on carbon to adatom tunneling parameters.
Adding the  contributions from the three first neighbors of the underneath C atom,   we get the following contribution to the low
energy Hamiltonian, 
\begin{equation}
< \Psi _{A,s,\sigma} |V_{T_{A(B)}}| \Psi _{ B,s,-\sigma } >=\frac 1 N  3   t _R \, s \, \frac {( 1 + \sigma \, s)} 2 \, \, .
\label{Top-Ras}
\end{equation}
This Rashba like SOC has the same form and sign  independently on the sublattice where the adatom is placed. The Rashba term gets its origin in the broken mirror symmetry produced by the adatoms and this is reflected  in  that $t_R$ change sign depending whether the top adatoms are  located on top or bottom of the graphene layer.


\section{Numerical results.}
Adatoms deposited on graphene should be placed at minimum energy equilibrium positions. The adsorption geometry depends on the particular heavy adatom\cite{weeks_2011}, being one of the  more interesting that in which the  adatoms place in  hollow positions.  When the adatoms are intercalated between graphene and the substrate, the adatoms form a superlattice commensurated with the graphene honeycomb lattice\cite{Calleja-2015}. It is also plausible to expect that low energy injected  adatoms become deposited in random positions. 

In this Section we show numerical results of the electronic structure  of graphene doped with heavy outer shell $p$-orbitals adatoms in three particular cases, i) the adatoms are randomly distributed in hollow positions, ii) the adatoms form a commensurate supercell with the graphene lattice and iii) the adatoms are fully random distributed on the graphene sheet. 

In the numerical calculations we consider a periodic rectangular graphene supercell of dimensions $L_x$=$N_x \sqrt{3}a$ and
$L_y$=$N_y a$,  defined by the lattice vectors, \cal {\bf A}=$N_y {\bf a}$ and \cal {\bf B} = $N_x (2 {\bf b} - {\bf a})$. In these expressions $N_x$ and $N_y$ are integer numbers. The unit cell contains 4$N_xN_y$ carbon atoms located at the graphene lattice positions $ \{ {\bf R}_i+{\bf d} _{\alpha} \}$. 
The adatoms are located at positions $\{ {\bf r} _i \}$. The concentration  of adatoms, $x$,  is given by the ratio  of number of  adatoms to the number of carbon atoms.

The electronic structure is obtained by diagonalizing the Hamiltonian,
\begin{equation}
H=H_0+\sum _{i,j,\sigma,\sigma'} \gamma _{i \sigma, j\sigma '} \left ( |Z,i,\sigma><Z,j,\sigma ' | + H.c. \right )
\label{Htotal}
\end{equation}
where $H_0$ is the pristine graphene Hamiltonian, Eq.\ref{TB-Graphene}, and  the second term describes the adatom induced hopping between carbon atoms. Because  of the  periodic boundary conditions, the electronic structure is described using the Bloch's theorem being the  electronic states characterized by a band index and  wavevectors $k_x$ and $k_y$ that are  defined in the interval
$[-\frac {\pi}{L_x},\frac {\pi}{L_x} ]$ and $[-\frac {\pi}{L_y},\frac {\pi}{L_y} ]$ respectively.  
In this geometry and for $N_y$  not being a multiple of three, the Dirac cones occur at  wavevectors
$\bf K$=$(0,\frac {2 \pi}{3L_y})$ and $\bf K ' $=$(0,\frac {4 \pi}{3L_y})$. For $N_y$ multiple of three the two Dirac cones overlap at the $\Gamma$ point. This overlap does not imply coupling between electronic states in different Dirac cones.  In order to simplify the analysis of the results, in this work we always consider supercells with  no overlapping Dirac cones.

Recent experiments seems to indicate that Pb on graphene can induce a large SOC, therefore in the numerical calculations we chose  Pb as the adatom,  and we use the tight-binding parameters obtained in reference\cite{Calleja-2015} for Pb atoms on graphene, $h$=0.27$nm$, $V_{pp\sigma}(h)$=0.4$eV$, $V_{pp\pi}(h)$=-0.6$eV$, 
$\epsilon _x$=$\epsilon_y$=1.65$eV$, $\epsilon _z$=1.38$eV$, $\Delta _{SO}$=0.9$eV$ and
$t$=2.7$eV$. 

 \begin{figure}[htbp]
 \includegraphics[clip,width=7.cm]{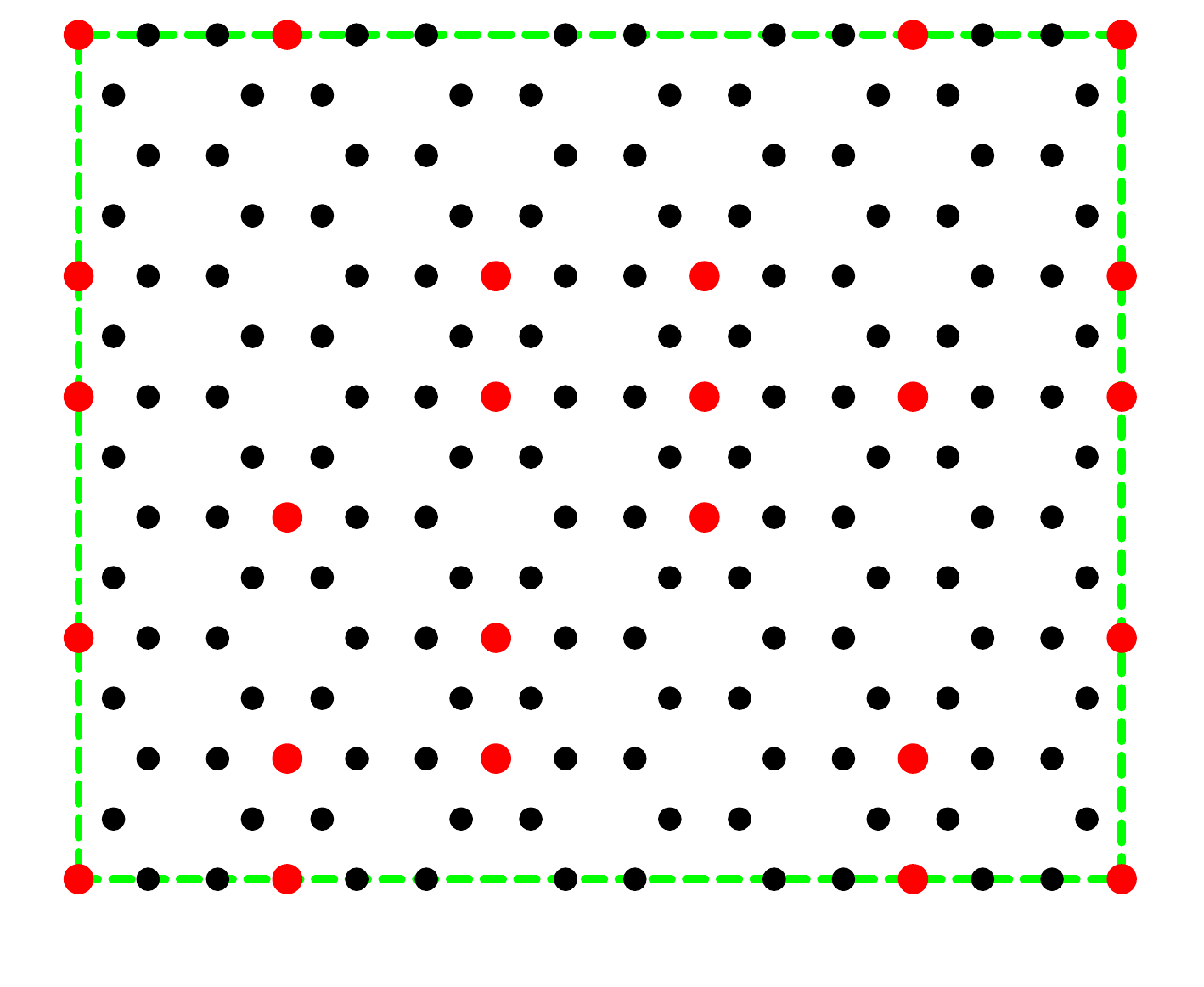}
\caption{(Color online)Graphene supercell with $N_x$=5 and $N_y$=7. Black small circles represent carbon atoms.
Larger red circles indicate the position of the adatoms. The adatoms are located in hollow positions and in this figure we plot a particular random realization of 
disorder. In this figure the number of adatoms per carbon atoms is $x$=17/140.
}                                     
\label{Figure3}
\end{figure}

\begin{figure}[htbp]
\includegraphics[width=\columnwidth,clip]{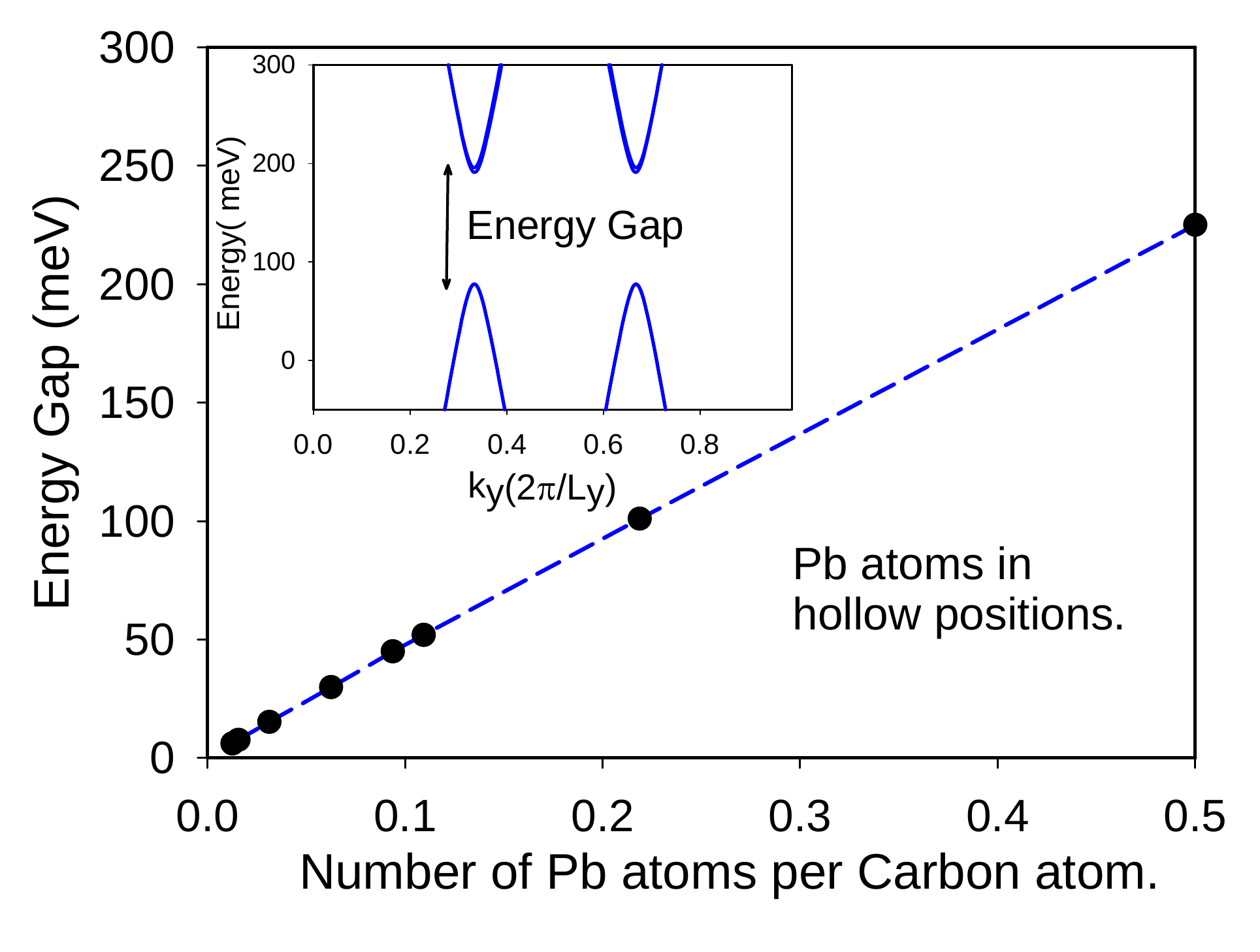}
\caption{(Color online) In the inset we plot  a band structure of a rectangular graphene supercell with $x$=0.25 Pb atoms per carbon. 
The Pb adatoms induce a gap at the Dirac points  ($0,\frac{2 \pi} {3 L_x}$) and  ($0,\frac{4 \pi} {3 L_x}$). As commented in the text,  the band structure is practically independent of the supercell size and the gap only depends on the concentration of Pb atoms. In the main figure we plot the gap  at the Dirac points as function of the Pb atoms concentration. The parameters used in the calculations are taken from reference \cite{Calleja-2015}.
}                                     
\label{Figure4}
\end{figure}

\subsection{Adatoms in Hollow positions.}
In this subsection we analyze  supercells with different sizes and forms and with different concentrations of adatoms randomly distributed, but always located in
hollow positions.
In Fig.\ref{Figure3} we show an example of supercell of size $N_x$=5, $N_y$=7 with 17 adatoms deposited in a random way in hollow positions.
In the inset of Fig.\ref{Figure4} we plot a typical band structure obtained for an adatom concentration $x$=0.25.  The adatoms open a gap at the Dirac points and,
in agreement with the results presented in subsection \ref{subsec:hollow}, the band structure corresponds to a Dirac equation in presence of an intrinsic SOC, equations \ref{Dirac} and \ref{SOcontIn}.
In our numerical calculations we obtain that for atoms adsorbed in hollow positions, the band structure always has this form, independently of  supercell size and  form  or  disorder realization.  The SOC  depends only  on the heavy atoms concentration.
In Fig.\ref{Figure4} we plot the energy gap as function of the adatom concentration, $x$. The dependence is practically linear and this indicates an  almost null interference effect between adatoms. 

We  understand this  linear dependence using Green function techniques. The low energy properties of an electron are described by the Dirac equation
$ \hbar v_F \sigma _0 \, ( s k_x \tau _x+ k_y \tau _y)$ 
and the corresponding Green function is
\begin{eqnarray}
G_0 ({\bf k},\omega)& =&  \frac 1 {\hbar \omega -H_0} \nonumber  \\
& = &  \frac {\sigma _0} {\hbar  \omega ^2 -\hbar   v_F ^2 k ^2}
\left ( 
\begin{array} {cc}  \omega &  v_F k e ^{i \theta _{\bf k}}  \\
 v_F k e ^{-i \theta _{\bf k}} &  \omega 
\end{array} 
\right )
\end{eqnarray}
Here $\theta _{\bf k}$=$\tan ^{-1} \frac {k_y}{s k_x}$. 
An adatom located in a hollow position, at ${\bf r}_i$, produces a scattering potential of the form, $s \, \sigma_z   \tau_z \delta (\bf r - \bf r _i)$.
In presence of a density, $x$, of adatoms in hollow positions and neglecting multiple scattering, the 
Green function of the total Hamiltonian  is\cite{Economou} 
\begin{equation}
G({\bf k},\omega)= G_0 ({\bf k},\omega) +   x  \Delta s  \sigma _z G_0 ({\bf k},\omega) \tau _z G ({\bf k},\omega) \, \, .
\end{equation}
Inverting this equation we get,
\begin{eqnarray}
G({\bf k},\omega)& =  & \frac 1 {\hbar ^2 \omega ^2 -\hbar ^2  v_F ^2 k ^2-  x^2 \Delta ^2  }  \times \nonumber \\
& &  \left  (   
\begin{array} {cc} \hbar \omega + x \Delta s \sigma _z  & \hbar v_F k e ^{i \theta _{\bf k}}  \\
\hbar v_F k e ^{-i \theta _{\bf k}} & \hbar \omega - x \Delta s \sigma _z
\end{array} 
\right )
\end{eqnarray}
that corresponds to the virtual crystal Hamiltonian
$H$=$H_0$+$x \Delta s  \sigma _z \tau_z$, describibg graphene in presence of  an intrinsic SOC of magnitude $x \Delta$.
\begin{figure}[htbp]
\includegraphics[width=\columnwidth,clip]{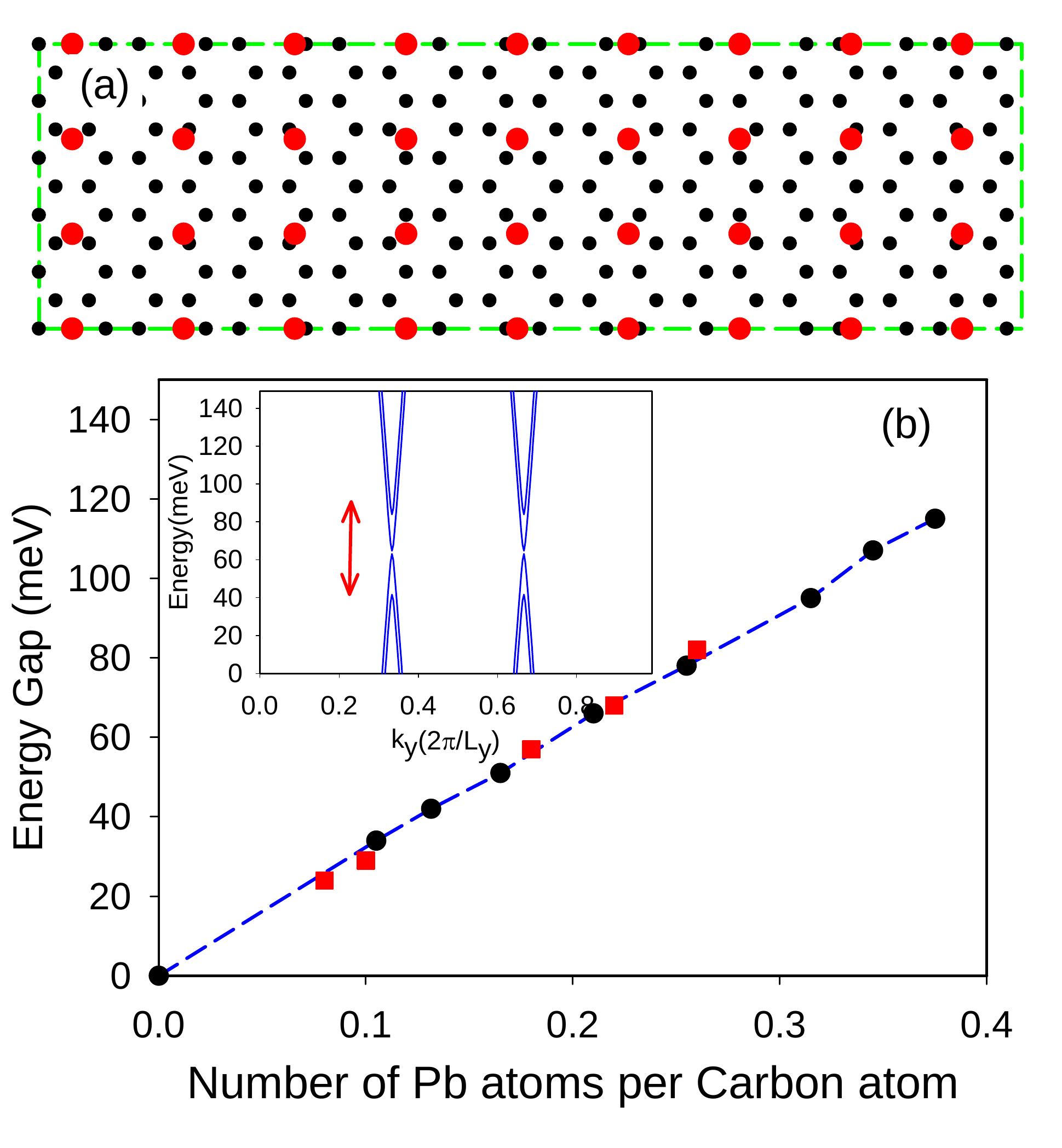} \caption{(Color online) (a) Graphene supercell with $L_x$=10$\sqrt{3}a$ and $L_y$=$5a$. Black small circles correspond to the positions of the carbon atoms.
Larger red circles represent adatoms. 
The adatoms form a rectangular lattice commensurate with the graphene supercell. In the figure the dimensions of 
the adatom cell are $l_x$=$L_x/9$ and $l_y$=$L_y/3$. (b) Inset, band structure for the geometry presented in (a). 
(b) Main figure, dependence of the energy gap, as defined in the inset, as function of the Pb atoms concentration. Black points are obtained in the graphene supercell
fixing the dimension $l_y$=$L_y/3$ and changing $l_x$. Red points are obtained fixing $l_x$=$L_x/9$ and changing $l_y$. The parameters used in the calculation are taken from reference \cite{Calleja-2015}. }                                     
\label{Figure5}
\end{figure}
\subsection{Commensurate array of Pb atoms on Graphene.}
Graphene grown on Ir(111)  forms a 9.3x9.3 moir\'e superstructure
with a $\sim$25{\AA} periodicity\cite{Diaye-2008}. When Pb atoms are intercalated under the graphene monolayer, the Pb atoms form a rectangular lattice commensurate with Ir. Therefore the honeycomb graphene lattice and  the array of Pb atoms  commensurate in a large moir\'e supercell\cite{Calleja-2015}.
 
In this subsection we study the spin-orbit effects induced by a rectangular  array of Pb atoms of dimensions
$l_x \times l_y$ commensurate with 
a  large rectangular graphene supercell of dimensions $L_x \times L_y$, see Fig.\ref{Figure5}(a). 
In the inset of Fig.\ref{Figure5}(b) we plot, for the geometry shown in Fig.\ref{Figure5}(a), the electronic states obtained with the tight-binding parameters corresponding to Pb. The band structure coincides with  the eigenvalues of the Dirac equation in presence of a Rashba like spin-orbit coupling. The intensity of the SO coupling is proportional to the energy gap between the second conduction band and the second valence band.
We find that this gap  increases linearly with the Pb concentration,$x$, and only depends on the concentration of Pb atoms being  independent of geometrical details. 
In Fig.\ref{Figure5}(b) we plot the energy gap as function of $x$ for a graphene supercell characterized by $N_x$=10 and $N_y$=5, and different combination of $i_x$ and $i_y$.
We obtain the same linear dependence for larger graphene supercells.

These results  indicate that Rashba coupling induced  from different adatoms do not interfere and the total Rashba coupling is just the sum of the different contributions. 
The Rashba SO coupling induced by Pb adatoms  have always the same sign, dictated
by the broken mirror symmetry, and the  linear behavior reveals  that in the commensurate phase, the adatoms average  all possible locations in the graphene unit cell.
On the contrary,  the absence of a gap at the Dirac points  indicates that the contribution to intrinsic SOC from  adatoms located in different places 
sums zero. 
This occurs because the sign of the intrinsic SOC induced by adatoms depends on its location.
A particular example is the case of the intrinsic SOC induced by adatoms in hollow position, Eq.\ref{Intri_H},  that has opposite sign than the induced by adatoms   located in top positions Eq.\ref{Intri_T}.

\begin{figure}[htbp]
\includegraphics[width=\columnwidth,clip]{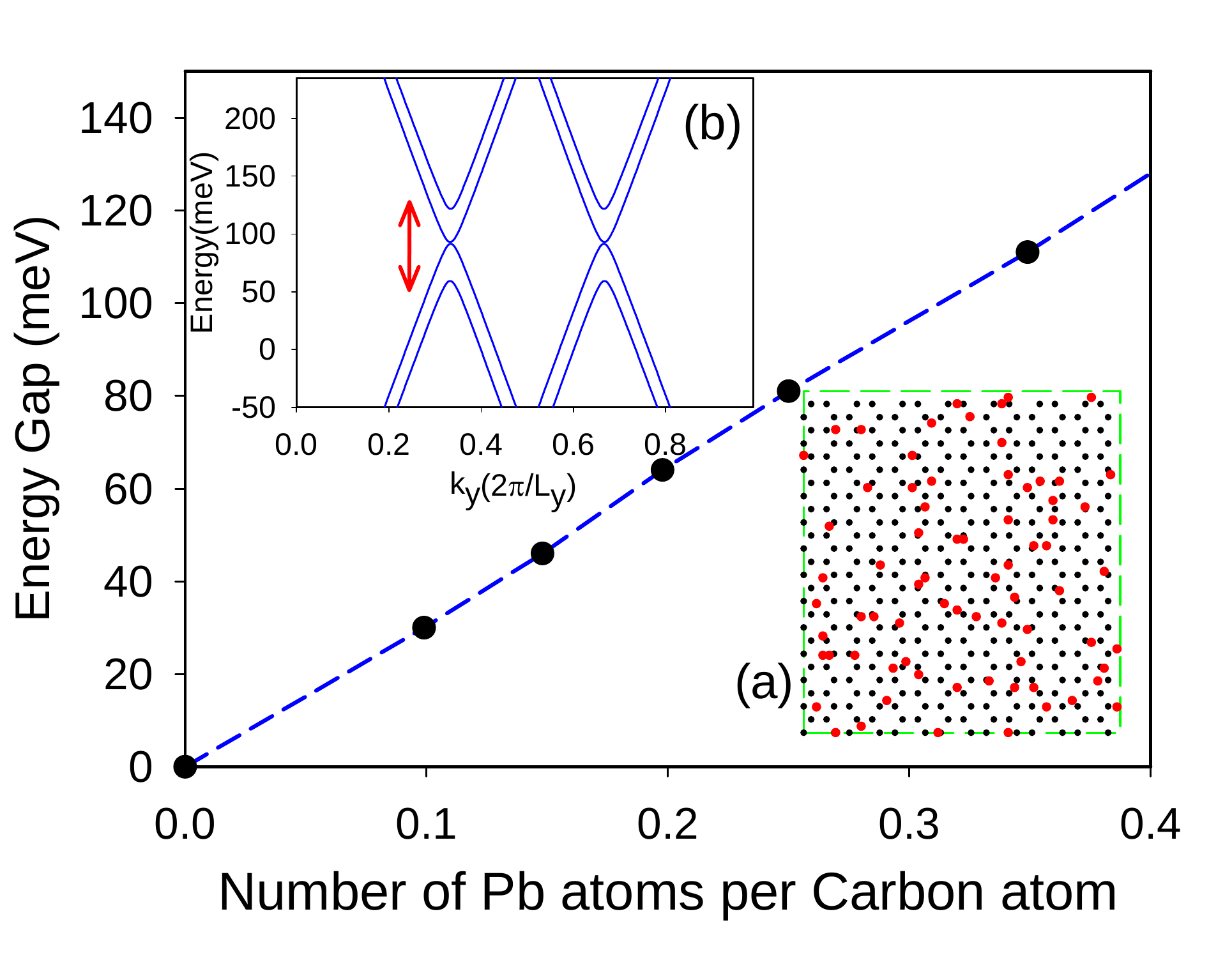}
\caption{(Color online) (a) Graphene supercell with $L_x$=7$\sqrt{3}a$ and $L_y$=$13a$. Black small circles correspond to the positions of the carbon atoms.
Larger red circles represent adatoms. 
The concentration of adatoms is $x$=0.2 and they are located in a random way.  (b) Inset, band structure for the geometry presented in (a). 
Main figure, dependence of the energy gap, as defined in the inset, as function of the Pb atoms concentration. The energy gap only depends on Pb concentration and is independent on geometry or disorder realization.  The parameters used in the calculation are taken from reference \cite{Calleja-2015}. }                                     
\label{Figure6}
\end{figure}

\subsection{Random Positions}

Finally we compute the SO induced  in graphene by a concentration of Pb adatoms randomly distributed. 
We study large graphene unit cells,  Fig.\ref{Figure6}(a), with different concentration of adatoms. The main results of the simulations are that there is not band gap at the Dirac points of the 
band structure, Fig.\ref{Figure6}(b), and the Rashba like SOC increases linearly with the concentration of Pb atoms, Fig.\ref{Figure6}. These results are independent on the disorder realization and size and form of the unit cell.  The dependence of the SOC on Pb concentration is the same than in the case of commensurate supercell. This, and the absence of intrinsic like SOC indicate that in both cases, commensurate order and random positions, there is not interference effects between adatoms, and the resulting SOC is just the sum of the contributions from  adatoms placed in different positions.

\begin{figure}[htbp]
\includegraphics[width=\columnwidth,clip]{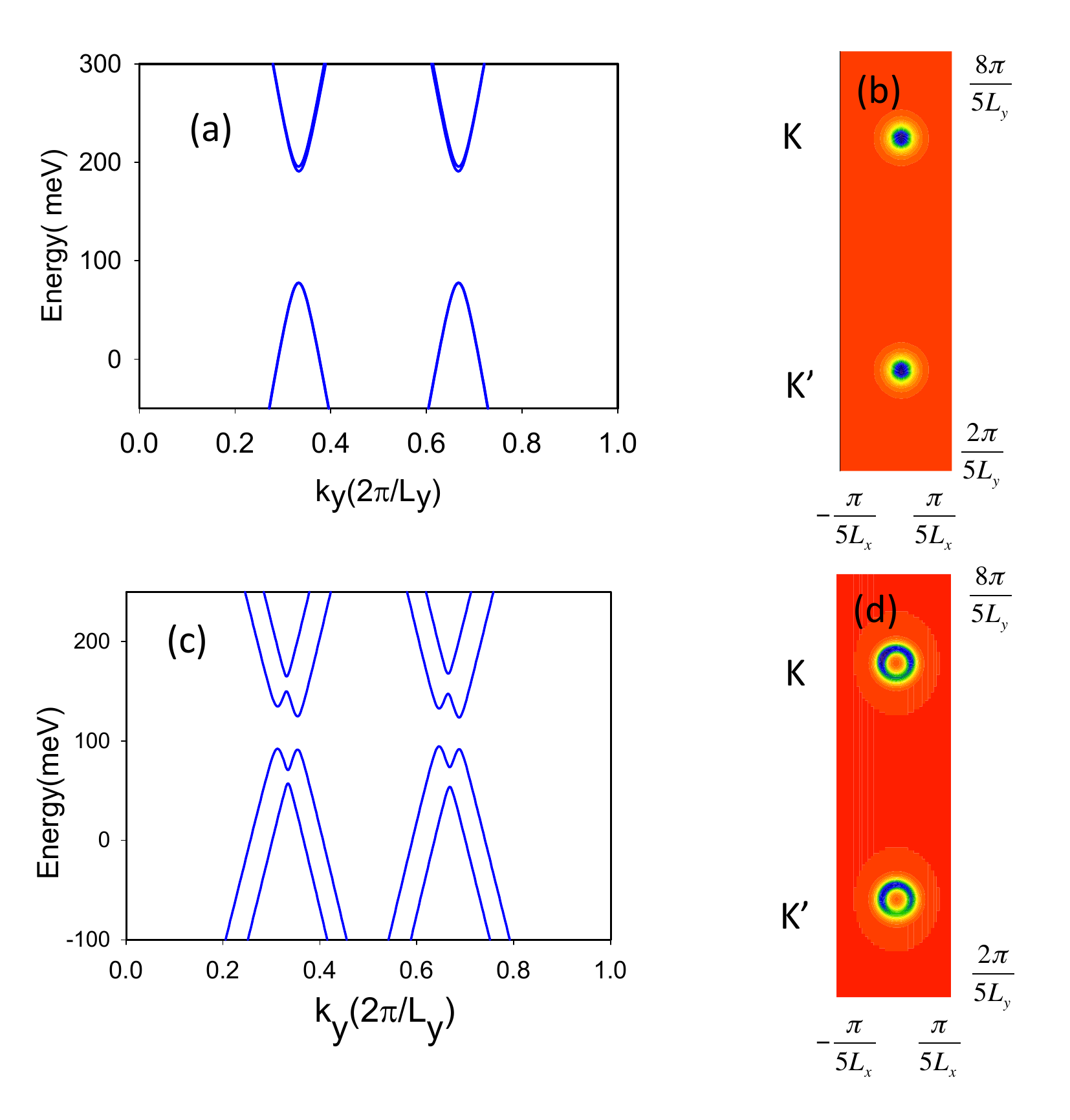}
\caption{(Color online)(a)Band structure of a graphene supercell ($N_x$=5, $N_y$=7) with a concentration $x$=0.25 of adatoms randomly located in hollow positions.
(b) Partial section of the Brillouin zone    showing in green, the regions that contribute to the spin  up Hall conductivity $\sigma _{xy} ^{\uparrow}$ for the case (a).
(c) Same than (a) for a fully random distribution of adatoms and an exchange field of 18$meV$. 
(d) Partial section of the Brillouin zone   showing in green the regions that contribute to the total Hall conductivity $\sigma _{xy} ^{\uparrow}$ + $\sigma _{xy} ^{ \downarrow}$ for the case (c).}                                     
\label{Figure7}
\end{figure}

\section{Topological Properties.}
In the previous Sections we have obtained that a graphene layer doped with adatoms placed in hollow positions has a gapped energy band structure similar to that  obtained from the Dirac equation with an intrinsic SOC. On the contrary  adatoms randomly or commensurately distributed on graphene  generate a gapless band structure  that remind that of  graphene with  Rashba SOC. In this Section we check that both adsorption geometries, hollow and random, have the same topological properties that the Dirac equation plus intrinsic and Rashba SOC respectively.  In order to know the topological properties, we compute the   spin resolved Hall conductivity of the system,
\begin{equation}
\sigma _{xy} ^{\sigma} = -2 \sum _{n,n',{\bf k}} \frac {{\rm Im} \left ( <n {\bf k}|v _{x} \hat P _{\sigma} | n' {\bf k} >< n' {\bf k}|v _y \hat P_{\sigma }|n {\bf k}>  \right ) } {(\varepsilon _{n, {\bf k} }- \varepsilon _{n',{\bf k}} ) ^2}
\end{equation}
where $|n {\bf k}>$ and $\varepsilon _{n {\bf k}}$ are the eigenfunction and eigenvectors respectively of the supercell Hamiltonian Eq.\ref{Htotal}, in the sum the index $n$ and $n ' $ run over occupied and empty states respectively, $v_\nu$=-$\frac 1 {\hbar } \frac {\partial H }{ \partial k_{\sigma}}$ is the velocity operator in the $\nu$-direction and $\hat P _{\sigma}$ projects the wave function in the subspace of spin $\sigma$.

In the case of adatoms in hollow positions we have obtained 
\begin{equation}
\sigma _{xy} ^{\sigma} = \sigma \frac {e^2} h \, \, \, \, \, \, \, \rm{adatoms \, in \, hollow \, positions.}
\nonumber
\end{equation}
that corresponds to a quantum spin Hall system\cite{Kane-2005}. The total Hall conductivity sums zero, as it should be in a system with time reversal symmetry. The main contributions to the Hall conductivity come from circular regions centered at the Dirac points, Fig.\ref{Figure7}(a)-(b).

Adatoms placed randomly on graphene do not generate a gap in the band structure, and the Hall conductivity is zero. However, in references  \cite{Qiao_2010} and \cite{Tse_2011}, it was proposed that an exchange field applied to graphene in presence of Rashba SOC should open a gap and the system would be a  non trivial insulator characterized  by an anomalous quantized Hall effect.  We have applied an uniform exchange field to the randomly doped graphene and we have obtained a gapped band structure, Fig.\ref{Figure7}(c) and a finite Hall
conductivity,
\begin{equation}
\sigma _{xy} = \sigma _{xy} ^{\uparrow}+ \sigma _{xy} ^{\downarrow} =2 \frac {e^2} h \, \, \, \, \, \, \, \rm{adatoms \, in \, random  \, positions.}
\nonumber
\end{equation} 
that proves that adatoms placed randomly on graphene generated a Rashba SOC.  In this case, and because the form of the bands Fig.\ref{Figure7}(c),  the main contributions to the Hall conductivity come from annulus  centered at the Dirac points, Fig.\ref{Figure7}(d).

The realization of the quantum spin Hall effect and quantum anomalous Hall effect for hollow and random adatoms respectively , shows the stability of these topological phases even for a non uniform distribution of the adatoms. 

\section{Summary.}
In this work we have studied the  spin orbit coupling induced in graphene by heavy adatoms with active electrons residing in $p$-orbitals. Depending on the location of the adatoms we find
different induced SOC's. Adatoms located in hollow positions open a gap at the Dirac points and induce an intrinsic like SOC. 
However adatoms randomly placed or commensurate with the graphene lattice  maintain the system gapless and induce a Rashba like SOC. The adatoms only perturb the pristine graphene band structure near the Dirac points. 

We find that the SOC induced by the adatoms is additive and there is not interference effects or multiple scattering.  The topological properties of graphene with hollow or random adatoms are the same than those of the Dirac Hamiltonian in presence of intrinsic or Rashba SOC respectively.  The finite value of the Hall conductivity of  graphene doped in  different geometries  indicates the robustness of the topological phases against a non uniform distribution of the spin orbit coupling.

\begin{acknowledgments} This work has been supported by MEC-Spain under grant
FIS2012-33521. 

\end{acknowledgments} 

%

\end{document}